\pdfoutput=1 
\documentclass{PoS}
\usepackage[authoryear]{natbib}
\bibpunct{(}{)}{;}{a}{}{,}

\title{SKA synergy with Microwave Background studies}

\ShortTitle{SKA synergy with Microwave Background studies}

\author{
\speaker
{Carlo Burigana},$^{ab}$
{Paul Alexander}$,^{c}$
{Carlo Baccigalupi}$,^{d}$
{Domingos Barbosa}$,^{e}$
{Alain Blanchard}$,^{f}$
{Adriano De Rosa}$,^{a}$
{Gianfranco de Zotti}$,^{gd}$
{Fabio Finelli}$,^{a}$
{Alessandro Gruppuso}$,^{a}$
{Michael Jones}$,^{h}$
{Sabino Matarrese}$,^{i}$
{Alessandro Melchiorri}$,^{l}$
{Diego Molinari}$,^{j}$
{Mattia Negrello}$,^{g}$
{Daniela Paoletti}$,^{a}$
{Francesca Perrotta}$,^{d}$
{Roberto Scaramella}$,^{k}$
{and Tiziana Trombetti}$^{a}$
\\
        \llap{$^a$}INAF-IASF Bologna, Via Piero Gobetti 101, I-40129, Bologna, Italy \\
        \llap{$^b$}Dipartimento di Fisica e Scienze della Terra, Universit\`a degli Studi di Ferrara, Via Giuseppe Saragat 1, I-44100 Ferrara, Italy \\
        \llap{$^c$}Cavendish Laboratory, Cambridge, UK \\
        \llap{$^d$}SISSA, Astrophysics Sector, Via Bonomea 265, 34136 Trieste, Italy\\
        \llap{$^e$}Radioastronomy Group, Instituto de Telecomunica\c c\~oes, Campus Universitario de Aveiro, 3810-183 Aveiro, Portugal \\
        \llap{$^f$}Universit\'e de Toulouse, UPS-OMP, IRAP, Toulouse, France \\
        \llap{$^g$}INAF -- Osservatorio Astronomico di Padova, Vicolo dellÕOsservatorio 5, Padova, Italy \\
        \llap{$^h$}Department of Physics, Oxford University, Oxford, UK \\
        \llap{$^i$}Dipartimento di Fisica e Astronomia G. Galilei, Universit\`a degli Studi di Padova, Via Marzolo 8, 35131 Padova, Italy \\
        \llap{$^l$}Dipartimento di Fisica, Universit\`a La Sapienza, P. le A. Moro 2, Roma, Italy \\
        \llap{$^j$}Instituto de F\'{i}sica de Cantabria (CSIC-Universidad de Cantabria), Avda. los Castros s/n, Santander, Spain \\
        \llap{$^k$}INAF -- Osservatorio Astronomico di Roma, Via Frascati 33, 00040 Monteporzio Catone (Roma), Italy \\             

        E-mail: \email{burigana@iasfbo.inaf.it}
}

\abstract{
The extremely high sensitivity and resolution of the Square Kilometre Array
(SKA) will be useful for addressing a wide set of themes relevant for
cosmology, in synergy with current and future cosmic microwave background
(CMB) projects.  Many of these themes  also have a link with future
optical-IR and X-ray observations.  We discuss the scientific perspectives
for these goals, the instrumental requirements and the observational and
data analysis approaches, and  identify several topics that are important
for cosmology and astrophysics at different cosmic epochs.
}

\FullConference{
Advancing Astrophysics with the Square Kilometre Array\\
June 8-13, 2014\\
Giardini Naxos, Italy}

\def\lsim{\,\lower2truept\hbox{${< \atop\hbox{\raise4truept\hbox{$\sim$}}}$}\,}
\def\gsim{\,\lower2truept\hbox{${> \atop\hbox{\raise4truept\hbox{$\sim$}}}$}\,}

\begin{document}

\section{Introduction}
\label{sec:Intro}

\vskip -0.3cm
Although it is not specifically designed for cosmic microwave background
(CMB) observations, because of its high resolution and limited
high-frequency coverage, the extremely high sensitivity
of the SKA (\citealt{dewdneyetal2013}, \citealt{braun2014})
may be used to address a wide set of themes relevant for
cosmology, in synergy with current and future CMB projects.  
Many of these  also have a strong link with 
the future {\it Euclid}\footnote{www.rssd.esa.int/euclid} mission 
and {\it Athena}\footnote{http://www.the-athena-x-ray-observatory.eu/} observations (\citealt{takahashi etal2014thisissue}).

In this chapter we will discuss several important themes:
(i) the contribution to future high-precision CMB absolute temperature experiments aimed at detecting spectral distortion, including long wavelength free-free emission 
linked to cosmological reionization (Sect. \ref{sec:cmb_spect});
(ii) cross-correlation between CMB and SKA surveys for the analysis of
Integrated Sachs-Wolfe (ISW) effect and constraining on dark energy (Sect. \ref{sec:CrossCMBradiosrc});
(iii) constraining non-Gaussianity with joint analyses of CMB and
radiosources (Sect. \ref{sec:CrossCMBradiosrc}); 
(iv) primordial
magnetic fields (Sect. \ref{sec:PMF});
(v) Galactic foreground
studies, linked to component separation in future CMB experiments
(Sect. \ref{sec:GalFore}).

From the observational point of view, these topics can be divided into two broad categories (\citealt{buriganaetalSKA2004}): 
those relying on precise point source observations and those that require sensitivity to extended structures. The former take
advantage of the high resolution and point-source sensitivity
naturally offered by interferometry, while the
latter require a suitable array design and observational approach,
such as a compact configuration implementing short baselines and a high
$uv$-space filling factor, mosaicing techniques, and for the largest
areas, methods such as the on-the-fly-mapping. These approaches, required
by a wide set of SKA scientific projects, are still under definition
and optimization.\footnote{See
  http://www-astro.physics.ox.ac.uk/~hrk/SKA$\_$EXPOSURE.html for a
  numerical tool.}
We will identify the corresponding instrumental and observational requirements and data analysis approaches
 needed to carry out the proposed studies.


\section{SKA contribution to future CMB spectrum experiments}
\label{sec:cmb_spect}

\vskip -0.3cm
The current limits on CMB spectral distortions and the constraints on
energy dissipation processes in the plasma (\citealt{SB02}) ($|\Delta
\varepsilon / \varepsilon_i| \lsim 10^{-4}$ ) are mainly set by the
COsmic Background
Explorer\footnote{http://lambda.gsfc.nasa.gov/product/cobe/} (COBE) /
Far InfraRed Absolute Spectrophotometer (FIRAS) experiment
(\citealt{mather90}, \citealt{fixsen96}).  High accuracy CMB spectrum experiments from
space, such as the Diffuse Microwave Emission Survey (DIMES)
(\citealt{KOG96}, \citealt{KOG03}) 
at $\lambda \gsim 1$~cm and
FIRAS~II (\citealt{FM02}) at $\lambda \lsim 1$~cm, have been proposed to
constrain (or probably detect) energy exchanges 10--100 times smaller
than the FIRAS upper limits.  Experiments such as DIMES may probe
dissipation processes at early times ($z \gsim 10^5$) resulting in
Bose-Einstein-like distortions (\citealt{SZ70}, \citealt{DD80}, \citealt{BDD91a}) and free-free
distortions (\citealt{bart_stebb_1991}) possibly generated by heating
mechanisms at late epochs ($z \lsim 10^4$), before or after the
recombination era (\citealt{BS03a}) (or possibly cooling processes, see \citealt{stebb_silk},
although these are disfavoured by Wilkinson Microwave Anisotropy
Probe\footnote{http://lambda.gsfc.nasa.gov/product/map/current/}
(WMAP) data).  These possibilities have been recently
reconsidered in the context of new CMB space missions such as the Primordial
Inflation Explorer (PIXIE) (\citealt{kogutpixie}) proposed to NASA,
which combines high-accuracy polarization and spectrum measurements at $\sim$
degree resolution, and  in the possible inclusion of spectral
measurements in highly sensitivite polarization
CMB space missions with arcmin resolution, such as the Cosmic Origins
Explorer\footnote{http://www.core-mission.org/} (COrE) (\citealt{corecoll})
proposed to the ESA
and its possible
successors, such as
the Polarized Radiation Imaging and Spectroscopy
Mission\footnote{http://www.prism-mission.org/} (PRISM)
(\citealt{prismjcap}).

\begin{figure}[t]
\minipage{0.45\textwidth}
\vskip -1.5cm
\includegraphics[width=\linewidth]{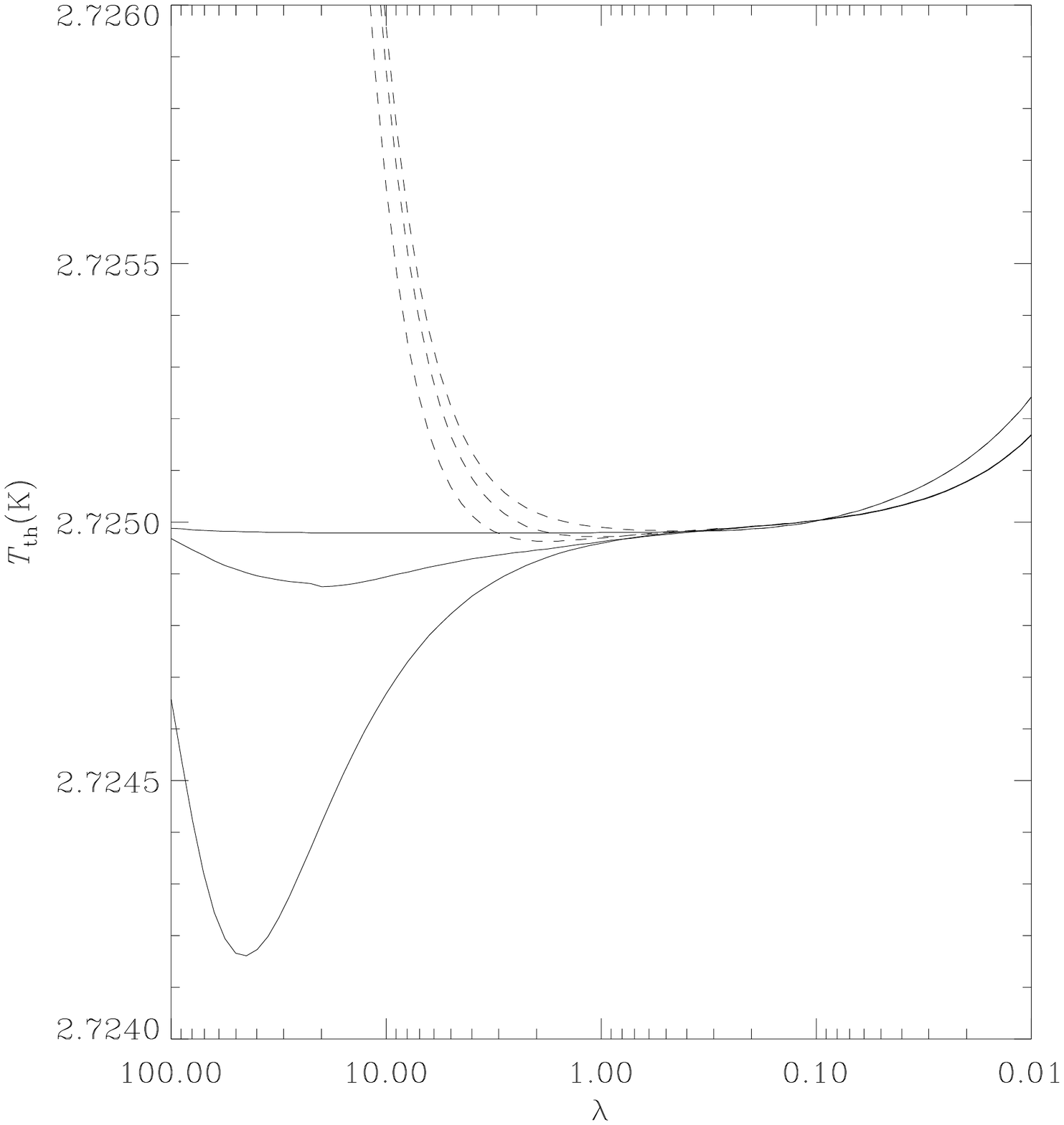}
\endminipage\hfill
\minipage{0.45\textwidth}
\vskip -1.5cm
\hskip -0.7cm
\includegraphics[width=\linewidth]{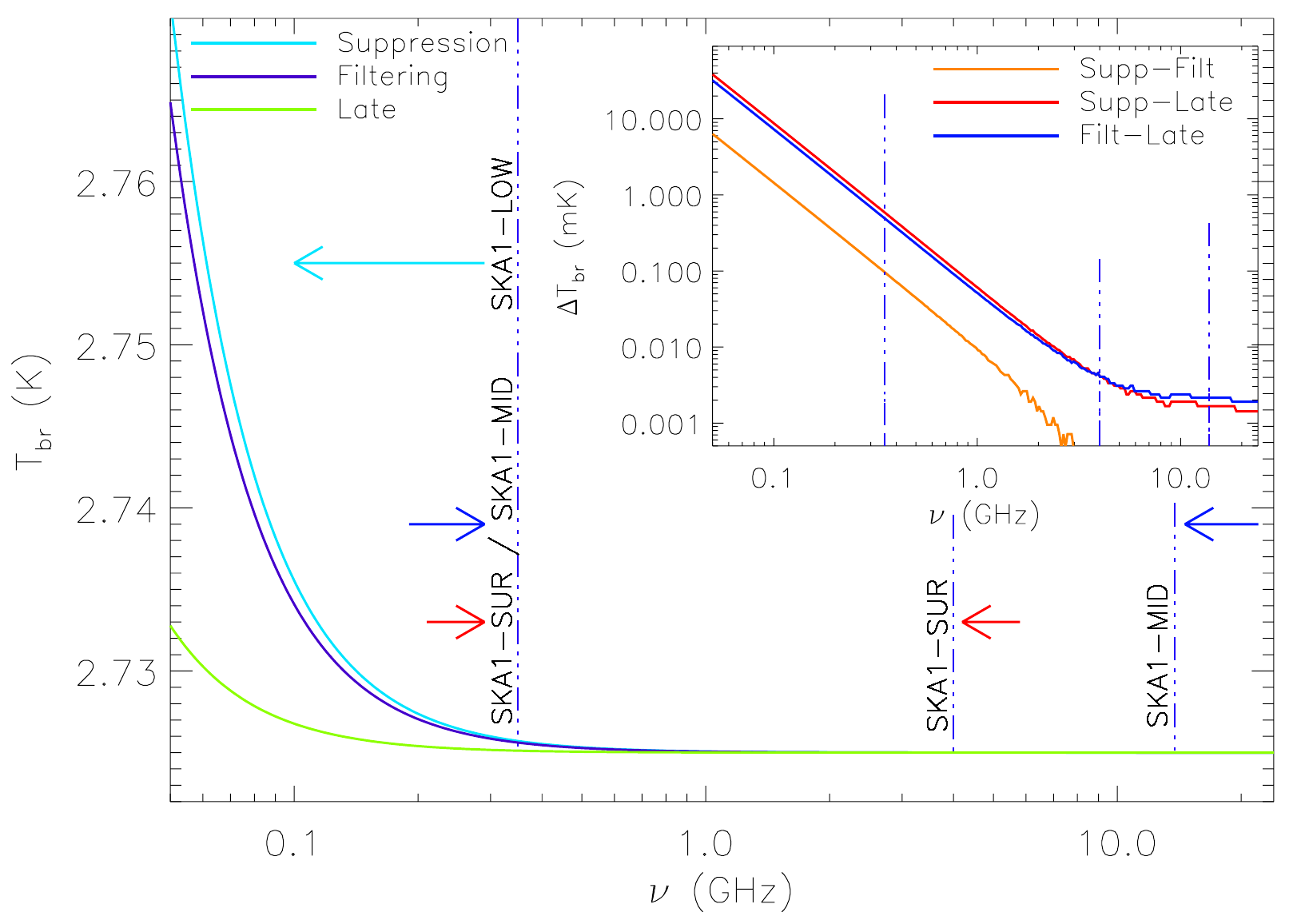}
\endminipage\hfill
\vskip -1.5cm
\caption{Left panel: CMB distorted spectra 
in terms of equivalent thermodynamic temperature
as a function of the wavelength
$\lambda$ (in cm) in the presence of
a late energy injection with $\Delta \varepsilon / \varepsilon_i \simeq 4y = 5 \times 
10^{-6}$ plus an early/intermediate
energy injection with $\Delta \varepsilon / \varepsilon_i = 5 \times 10^{-6}$
(about 20 times smaller than current upper limits)
occurring at the ``time'' Comptonization 
parameter
$y_h=5, 1, 0.01$ (from the bottom to the top;
in the figure the cases at $y_h=5$ 
-- when the relaxation to a Bose-Einstein modified spectrum 
with a dimensionless chemical potential given, in the limit of small distortions, by
$\mu \simeq 1.4 \Delta \varepsilon / \varepsilon_i$ is achieved -- 
and at $y_h=1$ are extremely similar at short wavelengths; solid lines) 
and plus a free-free distortion
with  $y_B=10^{-6}$ (dashes).
$y_h$ is defined by $y$
but with $T_e=T_{CMB}$ when the integral
is computed from the time of the energy injection
to the current time. From \citet{buriganaetalSKA2004}.
Right panel: free-free distortion 
in the SKA2 frequency range
produced by two astrophysical reionization histories 
(a reionization phenomenological model,
{\it late} model, see \citealt{naselskychiang04},
with parameters given by Eq. (4) of \citealt{trombettiburigana12} is also displayed for comparison).
The inset shows the absolute differences between the models. Vertical lines display the frequency ranges of the three SKA1 configurations.}
\label{fig:cmb_dist}
\end{figure}

Calculations of various types of typical distorted spectra (see left panel of Fig.~\ref{fig:cmb_dist})
have been presented in various works (e.g. 
\citealt{daneseburigana94}, \citealt{procopioburigana}, \citealt{khatrisunyaev}).
Improving CMB absolute
temperature measures will give a corresponding strengthening of the
constraints on physical parameters of various classes of processes (see e.g. \citealt{prismjcap} and references therein).
Decaying and annihilating particles during the pre-recombination epoch
may affect the CMB spectrum, with the exact distorted shape
depending on the process timescale and, in some cases, being different
from that produced by energy release.  This is especially interesting
for decaying particles with lifetimes $t_{X} \approx $\,few$\times 10^{8}
- 10^{11}$\,sec (\citealt{daneseburigana94}, \citealt{chlubasunyaev2012}).
Superconducting cosmic strings would also produce copious
electromagnetic radiation, creating CMB spectral distortion shapes (\citealt{ostrikerthompson87})
that would be distinguishable with high accuracy measurements.
Evaporating primordial black-holes provide another possible source of energy injection, with the shape of the resulting distortion depending on the black-hole mass function (\citealt{carretal2010}).
CMB spectral distortion measurements could also be used to constrain the spin of non-evaporating black-holes (\citealt{paniloeb2013}).
The CMB spectrum could also set constraints on the power spectrum of small-scale magnetic fields (\citealt{jedamziketal2000}),
the decay of vacuum energy density (\citealt{BartlettSilk1990}), axions (\citealt{ejllidolgov2014}), and other new physics processes. 
Deciphering these signals is a challenge, but holds the potential for important new discoveries and constraining unexplored processes that cannot be probed by other means.

In addition to the processes discussed above, a certain level of
departure of the CMB spectrum from a perfect blackbody is
theoretically predicted due to some unavoidable fundamental
processes. Cosmological reionization produces electron heating
which causes a Comptonization distortion proportional to the
fractional amount of energy exchanged during the interaction,
characterized by the Comptonization parameter
$y(t)=\int_{t_{i}}^{t}
[(\phi-\phi_{i})/\phi] (k_BT_e/m_ec^2) n_e \sigma_T c dt \simeq (1/4)
\Delta\varepsilon/\varepsilon_{i}$
 (where the last equality holds in
the limit of small energy injections and integrating over the relevant
epochs). Here $\phi (z) = T_{e}(z)/T_{CMB}(z)$, and $\phi_{i} = \phi
(z_{i}) = (1+ \Delta \epsilon/ \varepsilon_{i})^{-1/4} \simeq 1-y$ is
the  equilibrium matter temperature and radiation temperature ratio at the beginning of the heating process (i.e. at $z_{i}$).
Typical values of $y$ expected from reionization are $\approx 10^{-7}$.
For example, for two astrophysical reionization scenarios based on different radiative feedback assumptions
({\it filtering} and {\it suppression} models) \citet{buriganaetal08} found $y \simeq (0.965 - 1.69) \times 10^{-7}$.
Other kinds of unavoidable spectral distortions are Bose-Einstein (BE)-like distorted spectra, produced by the dissipation of primordial perturbations at small scales,
damped by
photon diffusion and thus invisible in the CMB anisotropies,
which produce a positive  chemical potential, $\mu_{0} \simeq 1.4
\Delta\varepsilon/\varepsilon_{i}$, and Bose-Einstein condensation of
the CMB by colder electrons 
associated with the faster decrease of matter temperature in the
expanding Universe relative to than that of radiation, 
which gives a negative chemical potential 
(for a recent review see \citealt{sunyaevkhatri2013}, and references therein). 
These two kinds of distortions are characterized by an amplitude, respectively, in the 
range
 $10^{-9} - 10^{-7}$
(and in particular $\simeq 2.52 \times 10^{-8}$ for a primordial
scalar perturbation spectral index $n_{S}=0.96$, without running), 
and $\simeq 3 \times 10^{-9}$. Since very small scales not explored by current CMB anisotropy data are relevant in this context, a wide range of primordial spectral indexes needs to be explored.
A wider range of chemical potentials is found by \citet{chlubaal12}, allowing also for variations of the amplitude of primordial perturbations at very small scales, as motivated by different inflation models. 

The free-free signal associated with cosmological reionization represents the most relevant type of low-frequency spectral distortion. 
The CMB brightness temperature, $T_{br}$, 
under the combined effect of Comptonization 
and free-free processes 
is approximated by
$(T_{br} - T_{CMB}\phi_{i})/T_{CMB} \simeq y_{B}/x^{2}-2y\phi_{i} \;,$
where $x=h\nu/kT_{CMB}$, 
and $y_{B}$ is the free-free distortion parameter. For a homogeneous ionized medium 
$y_{B} (x,t) = \int_{t_{i}}^{t}   (\phi-\phi_{i})\phi^{-3/2} g_{B}(x,\phi) K_{0B} dt$, where $g_{B}$ is the Gaunt factor weighted over ionized atoms
and the bremsstrahlung rate in a hydrogen-helium plasma is given by 
$K_{0B}(z) \simeq (8\pi/3) e^{6} h^{2} n_{e}^{free} (n_{H}^{+} + n_{He}^{+} +4 n_{He}^{++}) \phi^{7/2} / [m (6\pi m k T_{e})^{1/2} (kT_{e})^{3}]$
(\citealt{BDD95}).
The density of free electrons and ionized atoms is determined by the reionization history. 
Since structure formation is far from homogeneous, the dependence of free-free emission on the square of the baryon density implies a distortion amplification 
with respect to the case of homogeneous medium by a large factor, $\simeq 1 + \sigma^{2}$, 
where  $\sigma^{2}$ is the matter distribution variance
($\gg 1$ at moderate and low redshifts), as found in \citet{trombettiburigana14}
who combine Boltzmann codes for the matter variance evaluation with a
dedicated free-free distortion code.
The results are shown in right panel of Fig. \ref{fig:cmb_dist}, 
where both free-free signal and Comptonization decrement are included. 

The radio background, mainly contributed by the very bright Galactic synchrotron emission,
is approximated (in terms of brightness temperature) by $(T / {\rm K}) \sim 2.8 \times  (\nu/{\rm GHz})^{-2.55}$. 
The extragalactic background determined by the ARCADE 2 experiment (\citealt{seiffertetal2011}) (expressed in terms of equivalent  thermodynamic temperature) 
is also consistent with a power law with a very similar index,
$-2.57$, but with an amplitude about two times smaller,
plus a frequency-independent CMB contribution at $\simeq 2.725$\,K.
In order to accurately observe with dedicated experiments the tiny CMB spectral distortions discussed above, the problem 
of the modelling and subtraction of the contribution from 
Galactic emissions and extragalactic foreground needs to be
solved. The high sensitivity and resolution of the SKA can be used to address this issue, as discussed below. 

Current measurement of radio source counts at GHz frequencies 
(see e.g. \citealt{prandonietal01}, \citealt{condonetal2012})
have sensitivity levels of tens of $\mu \rm Jy$ (a recent estimation of
radio source background can be found in
\citealt{gervasietal2008}). However, the very faint tail of radio source
counts and their contribution to the radio background at very low
brightness temperature is not accurately known.
Exploiting the recent differential number counts
 at 0.153 GHz (\citealt{williamsetal2013}), 0.325 GHz (\citealt{mauchetal2013}), 1.4 GHz (\citealt{condonetal2012}), and 1.75 GHz (\citealt{vernstrometal2014}) 
it is possible to evaluate the contribution, $T_{b}$, to the radio background from extragalactic sources in various
ranges of flux densities. 
While these signals are clearly negligible compared to the accuracy of current 
CMB spectrum experiments, 
mostly at $\lambda \gsim 1$~cm,
they can be significant at the accuracy level 
potentially achievable with future experiments. 
Assuming to have subtracted the sources brighter than several tens of nJy,
$T_{b}$ is found to be less than $\sim 1$\,mK at frequencies above $\sim 1$\,GHz, but larger than $\sim 10$\,mK below $\sim 0.3$\,GHz.
The estimate of the minimum source detection threshold is given by  
the 
source confusion noise which, around 1.4\,GHz, has been quoted by 
 \citet{condonetal2012}:
$5\sigma_{conf} \simeq 5 \times 1.2 (\nu/3 \, {\rm GHz})^{-0.7} (\theta/8'')^{10/3} \mu$Jy, where $\theta$ defines the relevant resolution.
According to the authors,
the finite angular extension of faint galaxies, $\theta \sim 1''$, implies a ``natural confusion limit`` of about 10\,nJy at frequencies around $\sim 1.4$\,GHz, thus indicating that, 
for deep surveys such as those discussed below, source confusion will not represent a relevant limitation.

At $1\, {\rm GHz} \, \lsim \nu \lsim {\rm some \, \, GHz}$
($\lambda \approx 1$\,dm) the signal amplitudes found for CMB distorted spectra well below FIRAS constraints 
(see Fig.~\ref{fig:cmb_dist}) are significantly larger than the estimates of the background from extragalactic 
sources fainter than some tens of nJy. Free-free distortion amplitude increases at decreasing frequencies, but source confusion noise may represent there a serious problem, 
possibly preventing the achievement of the faint detection threshold necessary to have a source contribution to the background significantly 
less than the CMB distortion amplitude.

Extragalactic source contribution is small compared to the Galactic 
radio emission, which currently represents 
the major astrophysical problem in CMB spectrum experiments,
but, unlike the Galactic emission,
it cannot be subtracted from the CMB monopole 
temperature by exploiting its angular correlation properties. 
Accurate absolute measurements with a wide frequency coverage can allow
a joint fit of both CMB distorted spectra and astrophysical 
signals (see e.g. \citealt{SB02} for an application
to FIRAS data) but a direct radio background estimate
from precise number counts will certainly improve the robustness
of this kind of analyses. 

The relevance of this problem emerged in the detection by the 
Absolute Radiometer for Cosmology, Astrophysics, and Diffuse Emission
2\footnote{http://asd.gsfc.nasa.gov/archive/arcade/} (ARCADE 2) 
of an excess in the CMB absolute temperature at 3.3 GHz (\citealt{singaletal2011}, \citealt{seiffertetal2011}), 
which is likely explained by 
a background of very faint sources or by residual emission from subtraction of the Galactic component.
\citet{singaletalMN} concluded that,
for a radio background at the level reported, the majority of the total surface brightness would 
have to have been produced by ordinary star-forming galaxies at redshift $z \gsim 1$, characterized by an evolving radio--far-IR correlation, 
which changes towards the radio loud with increasing redshift. 
Using the JVLA at 3 GHz
\citet{vernstrometal2014mnras} ruled out a new discrete population peaking brighter than 50 nJy while \citet{vernstrometal2014} 
constrained the contribution to the ARCADE 2 excess from an extended population peaking above 1\,$\mu$Jy.
Analyzing dedicated radio observations with the JVLA,\footnote{http://www.nrao.edu/pr/2012/jansky/}
\citet{condonetal2012} argued that the discrete sources possibly dominating the reported extragalactic background excess
cannot be located in (or near) galaxies and must be typically fainter than 0.03 $\mu$Jy at 1.4 GHz.
The problem raised by this controversial 3.3 GHz excess underlines how crucial the precise estimation of very faint source counts is for the 
accurate exploitation of CMB spectrum measurements.

SKA continuum surveys, driven by a set of SKA top priority science cases and 
defined to provide significant advances over pre-SKA surveys, 
are described in 
\citet{PrandoniSeymour}. 
Deep and ultra-deep surveys are clearly the most relevant ones to determine the very faint source counts, while 
$P(D)$ methods can be exploited to extract information on number counts below the survey sensitivity (\citealt{condonetal2012}), particularly in 
low frequency continuum surveys, dedicated to non-thermal emission in clusters and filaments.
The Ultra Deep survey dedicated to the 
Star Formation History of the Universe (SFHU), with their planned rms sensitivity of some tens of nJy per beam 
and arcsec or better resolution, will represent a great opportunity
for an accurate determination of source number 
counts down to very faint fluxes, 
significantly helping the solution of one fundamental problem of the future generation of CMB 
spectrum experiments at long wavelengths.


\subsection{Free-free localized emissions}
\label{sec:ff_em}

\vskip -0.2cm
The SKA 
will be able, for the first time, to trace 
the detailed distribution of neutral hydrogen before reionization, and the neutral-to-ionized transition state
at the reionization epoch, through the 21-cm line (see
e.g. \citealt{Schneideretal2008}). It is also possible to trace the development of ionized material directly by looking for the free-free emission from ionized halos.  
The expected signal can be calculated
exploiting reionization phenomenological and astrophysical models through both semi-analytical
methods (\citealt{naselskychiang04},
\citealt{trombettiburigana14}), and numerical simulations (\citealt{ponenteetal2011}).

\begin{figure}[t]
\minipage{0.45\textwidth}
\vskip -0.7cm
\hskip 0.5cm
\includegraphics[width=\linewidth]{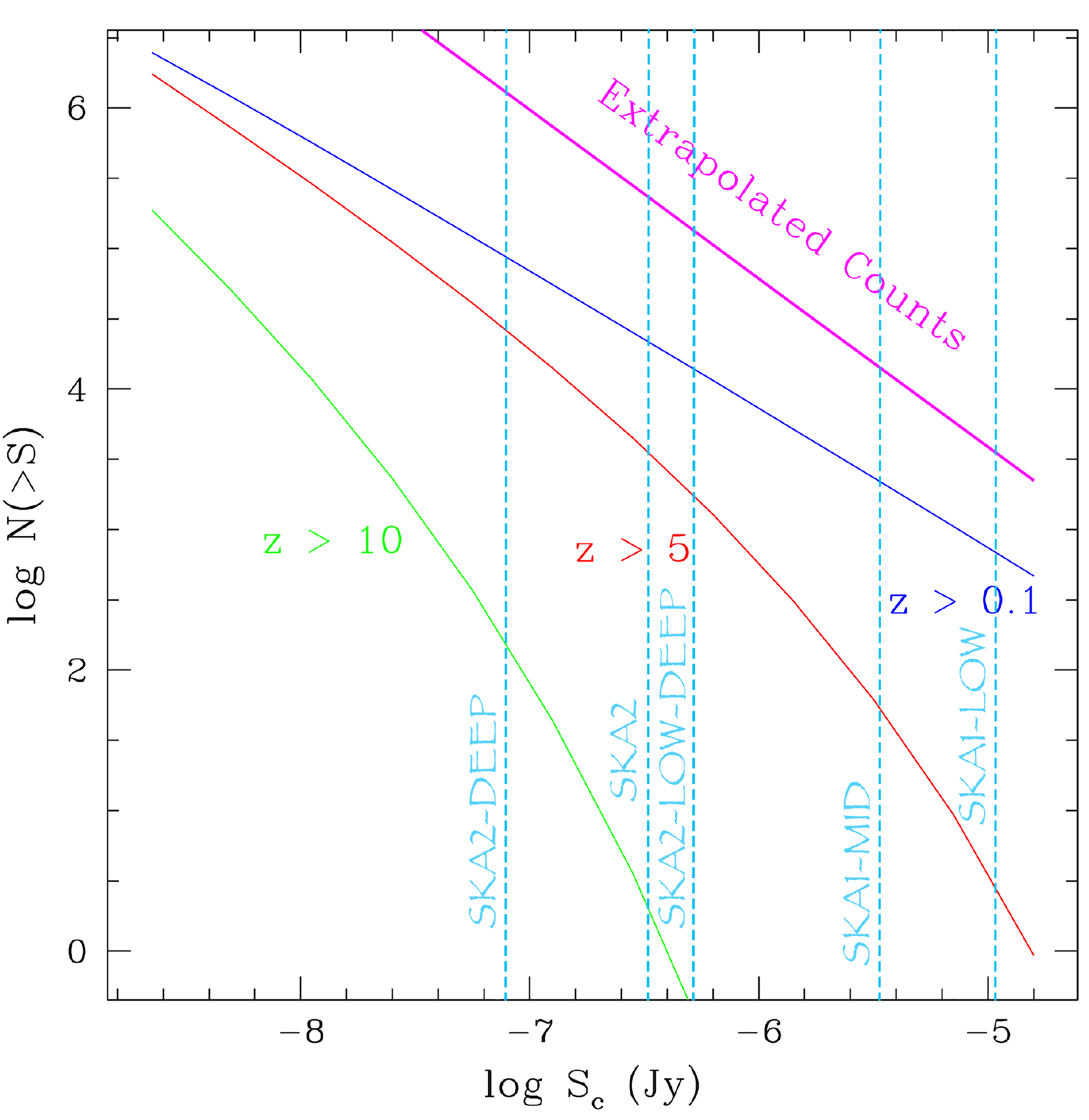}
\endminipage\hfill
\minipage{0.37\textwidth}
\vskip -1.2cm
\hskip -0.7cm
\includegraphics[width=\linewidth]{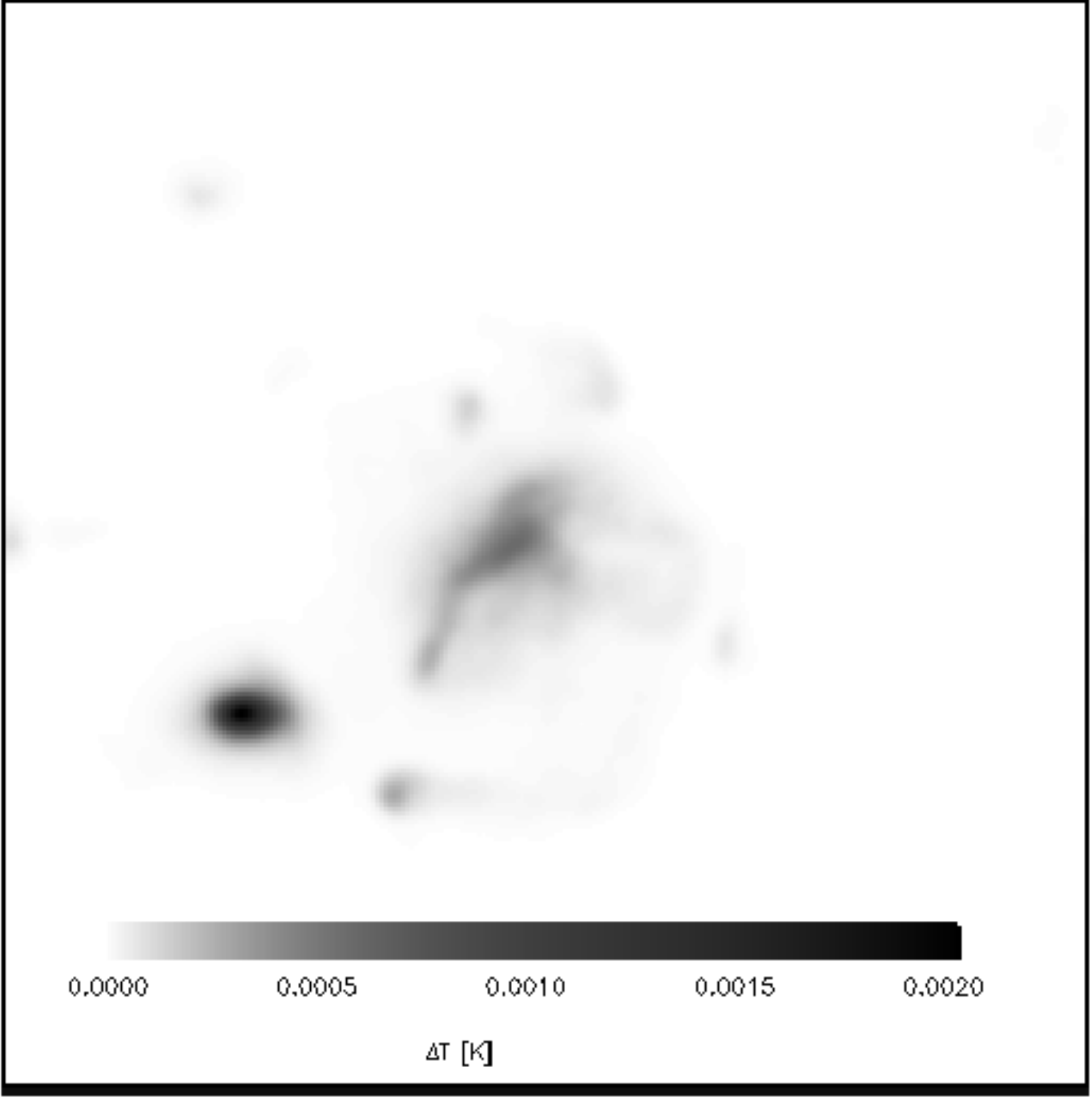}
\endminipage\hfill
\vskip -0.2cm
\caption{
Left panel:
number of sources which may be detected in 1 deg$^2$ by SKA, as a function of the threshold flux
$S_{c}$ (theoretical predictions from \citealt{OH99}). Realistic limiting fluxes (5$\sigma$ sensitivity levels
achievable in one hour of integration on one field of view (FoV))
for point source detection are shown (vertical lines). 
We report also the (5$\sigma$) sensitivity achievable by SKA2 with a deep exposure (in $\sim$ one day of integration) at low and mid/high frequencies.
Extrapolated source counts from \citet{Partridgeetal1997}.
Right panel: free-free signal for a 
halo with $M = 6.6 \times 10^{14} h^{-1} M_\odot$ at 
$z=0.15$. The greyscale shows the signal
in K and at 1 GHz. 
The field of view is $\approx 40'$. The total flux in this region is  $S_{\rm ff} = 2.83\times 10^{-5}$ Jy. From \citet{ponenteetal2011}.
}
\label{FigFF}
\end{figure}

The direct observation of diffuse gas and Pop III objects
in thermal bremsstrahlung has been investigated by \citet{OH99}. 
Observations at high resolution of dedicated sky areas are a natural way to distinguish free-free distortion by ionized halos rather than by diffuse ionized IGM.
The SKA should be able to detect up to $\sim 10^{4}$
individual free-free emission sources with $z>5$ 
in 1 square degree (see left panel of Fig. \ref{FigFF}).

Simulations by \citet{ponenteetal2011} show that the expected individual halo signal should
be $S_{\rm ff} \sim 3.67\times 10^{-9}$ Jy, while the analytic $\beta$-model for halo density profiles predicts a signal a factor $\sim 7.5$ larger. 
In terms of $\Delta T$, the maximum temperature distortion is about a few $\mu$K (at 1 GHz) at the center of the cluster. 
More massive and denser clusters would produce 
stronger signals, representing a useful way to study the intracluster medium. 
For a simulated massive cluster at redshift $z=0.15$ 
the free-free distortion at 1 GHz is 
of the order of 1 mK in the cluster regions (see right panel of Fig. \ref{FigFF}). 
Thus, the precise mapping of individual halos represents an interesting goal for the excellent imaging capabilities of the SKA.

\section{Cross-correlation with CMB}
\label{sec:CrossCMBradiosrc}

\vskip -0.3cm
High accuracy CMB surveys such as those carried out and expected by the {\it Planck} 
satellite\footnote{www.rssd.esa.int/planck}
are designed to cover 
high sky fractions, $f_{sky}$,
or the whole sky. 
The SKA is mainly designed to achieve very faint fluxes on limited sky fields, but its sensitivity is so high on typical FoVs of $\sim$ degree side at frequencies 
$\sim$ 1 GHz, that it is reasonable to expect to be able to cover a significant $f_{sky}$ (thousands of square degrees) 
with unprecedented sensitivity 
in some months of integration. This will improve the cross-correlation analyses with CMB surveys and with surveys in other frequency bands (see also \citealt{takahashi etal2014thisissue}). 

\subsection{Integrated Sachs-Wolfe effect and constraints on dark energy}
\label{subsec:ISWDE}

\vskip -0.3cm
The ISW effect comes from the line-of-sight integral in the Sachs Wolfe equation (\citealt{SachsWolfe}).
It arises when CMB photons streaming across the Universe interact with the time-evolving gravitational potential wells associated with foreground 
large-scale structure (LSS). The evolution of the potential leads to a net change of the photon energies as they pass through the LSS. The ISW is a linear effect 
which depends the cosmological model, since it requires  a change in the cosmic fluid equation of state. The evolution of the gravitational potential 
is related to the matter linear density perturbations; 
in the matter dominated regime, the growth of matter perturbation 
is proportional to the scale factor. This balances the dilution of matter due to the cosmic expansion and makes the gravitational potential 
variations negligible. They are relevant however at early times, when the Universe goes from being radiation dominated to matter dominated (\textit{early ISW}), 
and at late times, as the dark energy (DE) (or curvature) takes over from the matter (\textit{late ISW}). The ISW contribution to the CMB anisotropy  in a 
direction $\hat{n}$ on the sky is approximately given by 
$\Delta^{ISW}({\hat n}) \approx -2\int_{\rm Last \; Scattering}^{\rm Today} d\eta \  {\dot \Phi}[r\hat{n},\eta] \ ,$
where $\Phi$ is the Newtonian potential, the dot denoting a derivative with respect to the conformal time $\eta$, and $r(\eta)$ is the proper distance. 

Unlike the early ISW, the late ISW is virtually uncorrelated with the CMB anisotropies generated at last scattering.
Direct detection of late ISW is difficult because of its small amplitude and 
its dominance only on super-horizon (i.e. large) scales, where cosmic variance is large. However, it is possible to isolate the late ISW generated at low redshifts through the cross-correlation 
of the CMB maps with LSS surveys. Indeed, when CMB photons cross a time-varying potential, they become slightly hotter or colder: statistically, we expect a tiny correlation 
of hot spots in the CMB with LSS, an effect expected to be less than 1 $\mu$K, orders of magnitude smaller than the CMB correlations (\citealt{Crittenden}, \citealt{Peiris}). 
Several measurements have been performed to detect 
the ISW signal: positive cross-correlations were measured using Sloan Digital Sky Survey\footnote{www.sdss.org} (SDSS) galaxy data and WMAP 
(\citealt{Fosalba03}, \citealt{Padmanabhan}, \citealt{Granett08}, \citealt{Granett09}, \citealt{Papai}),
on APM galaxies (\citealt{Fosalba04}), 
the 2MASS survey ({\citealt{Afshordi}), and on radio data (see {\citealt{Nolta}, \citealt{Raccanelli} and \citealt{Boughn04a,Boughn04b}
where correlations with hard X-ray background were found). Also, 
\citet{Afshordi}, \citet{Rassat}, and \citet{Francis}
used IR galaxy samples to characterize the ISW signal. 
The typical significance of ISW detections is currently quite low, around 2-3 $\sigma$.
The cross-correlation detection of CMB with LSS requires a good CMB map on 
large scales and a deep enough galaxy distribution map with large $f_{sky}$ to reduce the 
uncorrelated CMB map noise, with $S/N \propto f_{sky}^{1/2}$.

Given a CMB map in temperature and a galaxy survey ${\bf x=(T,G)}$ (vector in pixel space), the quadratic maximum likelihood (QML) estimator (\citealt{Tegmark1997}) 
provides an estimate of the angular power spectrum (APS)
$\hat {C}_\ell^X$, where $X$ = $TT, TG, GG$.
QML is well suited for such analysis,
being optimal, i.e. unbiased and with minimum variance,
an essential feature when the S/N ratio is low, as for the ISW effect. 
It is computationally demanding, but can be applied at relatively low resolution, i.e. for large scales, where the (late) ISW effect appears,
and 
even for $f_{sky} <1$,
being a pixel based method.
A set of estimators (\citealt{Xia09})
exploiting the features of the redshift surveys considered 
has been developed:
the correlation of WMAP7 CMB data 
with radio sources in the NRAO Very Large Array Sky Survey\footnote{http://www.cv.nrao.edu/nvss/} (NVSS)
and its implications on the cosmological perturbation statistics (\citealt{Xia10a}); 
the foreground removal from CMB 
maps 
(\citealt{Xia10b})
to improve auto and cross-correlation spectra (\citealt{Xia11}); 
and WMAP7 maps and NVSS cross-correlations with dedicated 
methods to constrain the DE content (\citealt{schiavonetal2012}).

From the 2013
data release, the {\it Planck} Collaboration detected the ISW effect with a $2\,\sigma$ -- $4\,\sigma$ significance, depending on the adopted method (\citealt{planck19}).
The SKA will play a crucial role in improving the current large-scale galaxy dataset. 
The requirement to have a negligible shot noise in the galaxy power spectrum is $\beta \; \rm{FoV} \; T_0  > 10 \, {\rm deg}^2 {\rm yr} $ (\citealt{Abdalla}); here $\beta $ is the receiver bandwidth, $T_0$ is the survey duration and FoV is 1 deg$^2$ and 200 deg$^2$ for SKA1-MID and SKA2-LOW, respectively. It is foreseen that a 1-yr 
SKA survey will contain $ > 10^9 (f_{sky}/0.5)$ HI galaxies in a redshift range $0 < z < 1.5$. This makes the combination of SKA and {\it Planck} data a 
powerful tool for investigating the ISW correlation, offering the opportunity to achieve an independent measurement of the effect and increasing the confidence level in the detection from the present, marginal evidence. It will also be possible to improve constraints on the statistics of primordial perturbations and DE dynamics. 


\subsection{Non-Gaussianity from joint analyses of CMB and radiosources}
\label{subsec:NonGauss_CMBradiosrc}

\vskip -0.15cm
Statistical analyses of the extragalactic source distribution can probe the Gaussianity of primordial perturbations through their imprints at the origin of LSS. 
Different kinds of non-Gaussianity, such as the local type (loc), equilateral (eq), enfolded (enf), orthogonal, have been predicted (\citealt{Bartolo04}, \citealt{Komatsu10}). 
For example, the local type is parameterized by a constant dimensionless parameter $f_{NL}$ (see e.g. \citealt{Verde00}, \citealt{Komatsu01}, \citealt{Babich04}),
 $\Phi = \phi + f_{NL} ({\phi}^2-<{\phi}^2>)$, where $\Phi$ denotes
Bardeen's gauge-invariant potential (evaluated deep in the matter era in the CMB convention) and $\phi$ is a
Gaussian random field. Extragalactic radio sources are powerful tracers of the LSS, since they
span a large volume extending out to substantial redshift.  
A global analysis of the constraints on the amplitude of primordial non-Gaussianity (PNG) from the APS 
obtained from extragalactic radio sources (the SDSS Release Six quasar catalogue),  the final
SDSS II Luminous Red Galaxy (LRG) photometric redshift survey, and
cross-correlation with the WMAP CMB temperature map,  has set limits of $f^{loc}_{NL}=48\pm20$,
$f^{eq}_{NL} =50\pm265$ and $f^{enf}_{NL}=183\pm 95$ at 68\% confidence level, almost stable with respect to potential systematic errors and analysis details (\citealt{Xia11}). 
A recent, interesting analysis of PNG of local type using the clustering of $8\times10^{5}$
photometric quasars from the SDSS in the redshift range $0.5 < z < 3.5$ has been presented by 
\citet{Leistedt_etal2014}. The authors separate the sample of quasars into four redshift bins by selecting objects with photometric redshift estimates and mitigate the impact of systematics in the estimate of power spectra with a novel technique, the extended mode projection, based on the measurement and mapping onto the sky 
of the most potential systematics (e.g. observing conditions, calibration) during SDSS observations.
They obtain $-49 < f^{loc}_{NL} < 31$ and predict an error $\sigma (f^{loc}_{NL}) \simeq 5$ 
from the angular power spectra of galaxies in 20 tomographic bins in the redshift range 0.5 < z < 3.5 that will be obtained 
with the Large Synoptic Survey Telescope-like photometric survey (\citealt{LSSTDEColl}).
Furthermore, a robust constraint, $f^{loc}_{NL}=5\pm21$, 
on PNG of  local type has been derived by \citet{Giannantonio_etal}
cross-correlating 
a wide set of currently available catalogs of galaxy surveys and them with WMAP maps and
performing an extended analysis of the possible systematics aimed at reducing their impact on the results. 
Tests of non--Gaussianity would have profound implications for inflationary mechanisms -- such as single-field slow roll,
multifields, curvaton (local type) -- and for models whose effects on the halo clustering can be described by the
equilateral template (related to higher-order derivative type non-Gaussianity) and by the enfolded template
(related to modified initial state or higher-derivative interactions). 
{\it Planck} data already set strong limits  on the non-Gaussianity parameter $f_{NL}$, namely 
$f^{loc}_{NL}=2.7\pm5.8$, $f^{eq}_{NL} =-42\pm75$ and $f^{enf}_{NL}=-25\pm39$ at 68\% confidence level (\citealt{PlanckCollNonGauss}), 
significantly constraining or ruling out many classes of inflationary models. 

Recent forecast works exploit the SKA radio continuum surveys (\citealt{PrandoniSeymour}) covering $\simeq 3.1 \times 10^{4}$\,deg$^{2}$ out to high redshift.
To mitigate the problem of lack of redshift data, \citet{raccanellietal2014} propose a combination of two methods: 
cross-correlation of the radio counts with CMB temperature anisotropies, i.e. the ISW effect, 
to reduce systematics on large scales which are particularly sensitive to PNG; 
and cross-identification of radio sources with optical data in order to split the radio sources into redshift bins.
The authors show that even with only two redshift bins, a tomographic analysis could improve the constraints on $f^{loc}_{NL}$
by an order of magnitude with respect to the case of a single redshift bin, achieving $\sigma (f^{loc}_{NL}) \simeq 2$, while including full redshift information 
will allow highly precise measurements of the non-Gaussianity parameter, with $\sigma (f^{loc}_{NL}) < 1$.
General relativistic effects on the galaxy number counts, such as a non-linear primordial correction and 
linear projection effects from observing in redshift space on the past light-cone, have been included in the analysis of \citet{cameraetal2014}.
Their relevance emerges for a precise analysis of small values of $f^{loc}_{NL}$, necessary in the light of {\it Planck} results. 
Neglecting difficulties due to cosmological parameter degeneracies, since the inclusion of PNGs  
does not significantly broaden the constraints on other parameters, the standard approach is adopted
of varying freely only the parameters one is interested in and fixing all the others to their fiducial values, as fitted for instance by {\it Planck}. The authors find that with SKA2 (in its full configuration) it will possible to constrain $f^{loc}_{NL}$
down to $\sigma (f^{loc}_{NL}) \simeq 1.54$, thanks to the large number of HI galaxies that will be detected up to high redshift. These works indicate the possibility to improve with SKA the constraints on $f^{loc}_{NL}$ 
of a factor $\sim 3$ with respect to {\it Planck} results.

The combination of high accuracy and deep extragalactic source surveys achievable with the SKA and the forthcoming and future
 CMB maps will provide a significant progress in this topic, at least for the local configuration.


\begin{figure}[t]
\minipage{0.5\textwidth}
\vskip -1.cm
\hskip -0.3cm
\includegraphics[width=\linewidth]{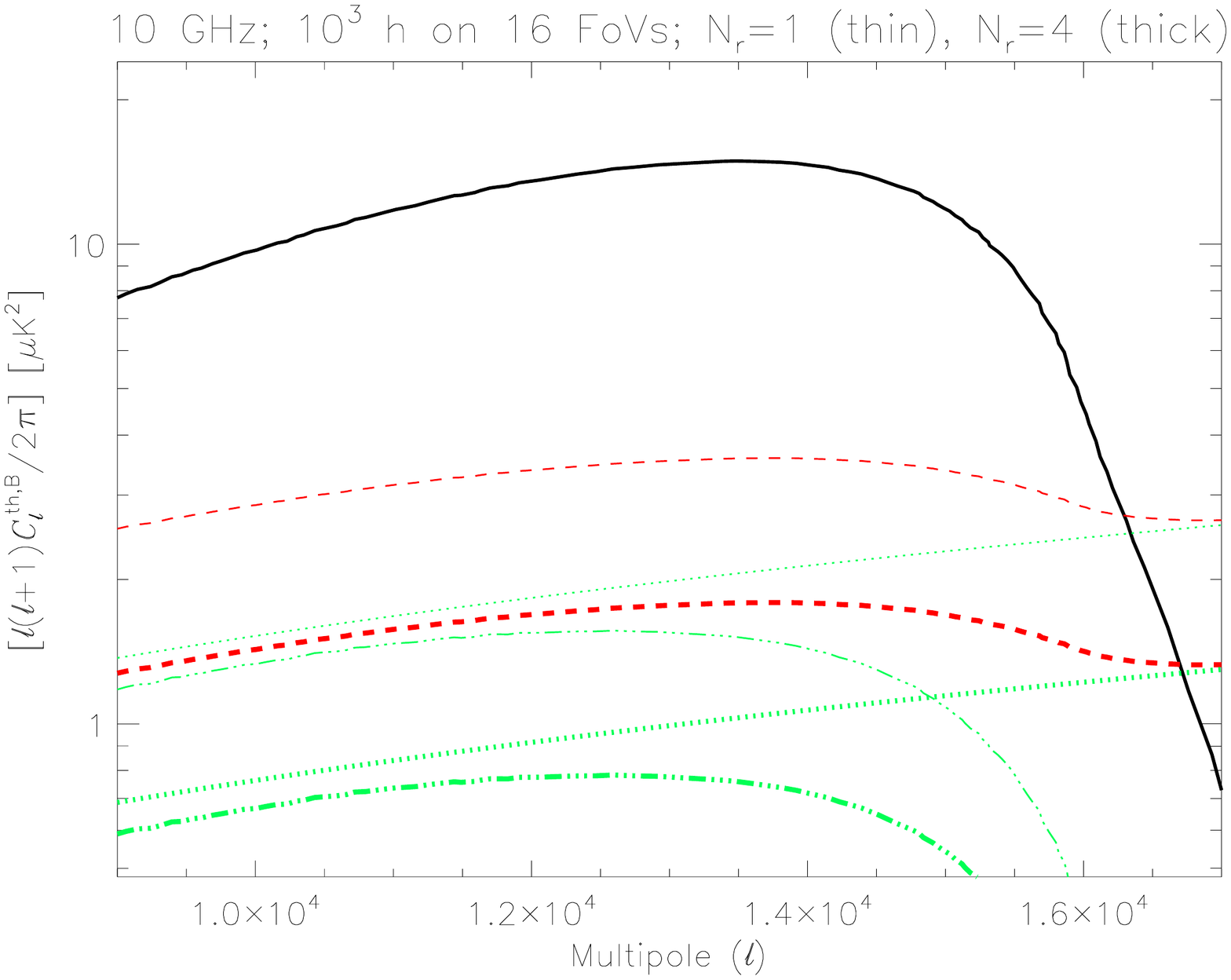}
\endminipage\hfill
\minipage{0.5\textwidth}
\vskip -1.cm
\hskip -0.5cm
\includegraphics[width=\linewidth]{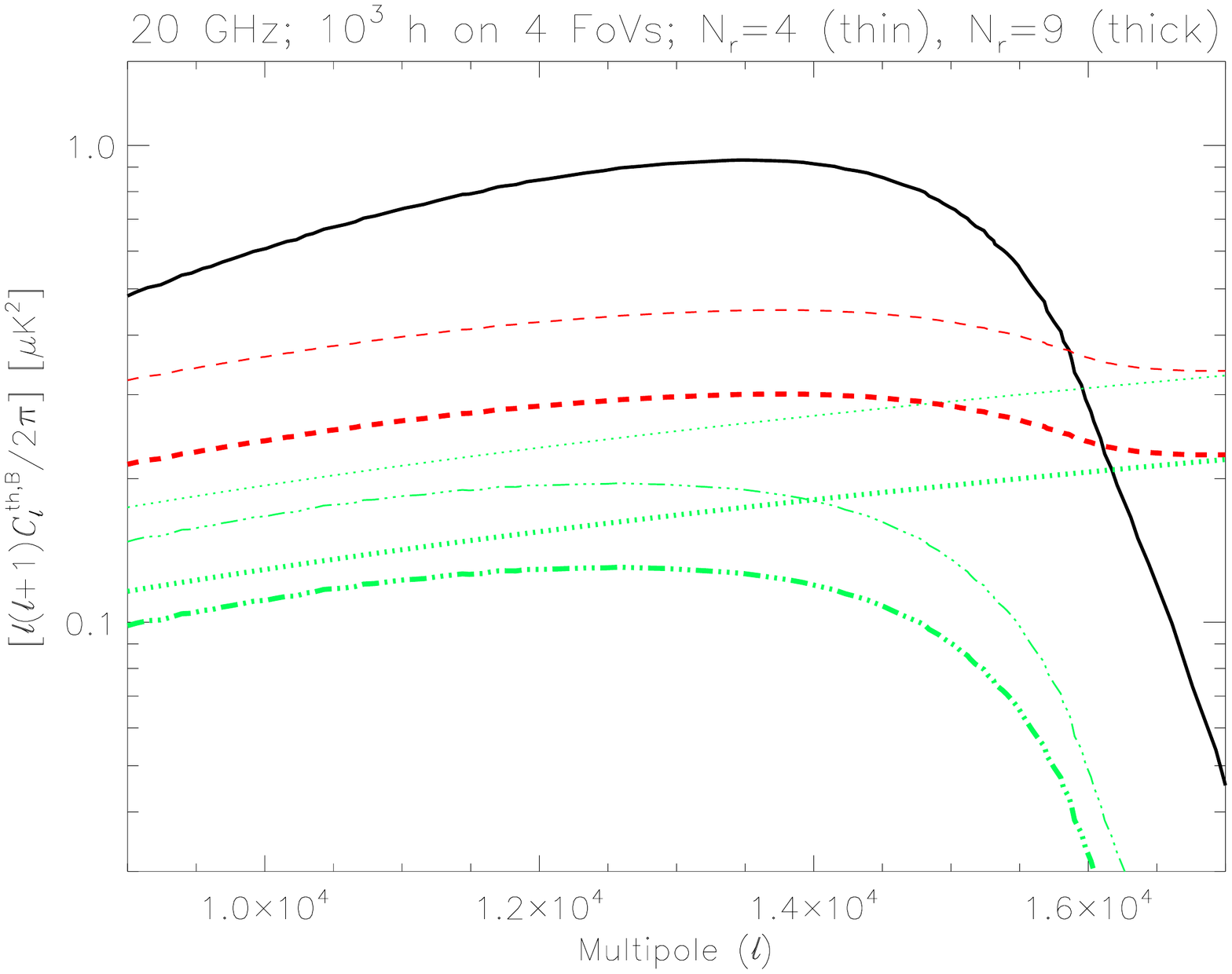}
\endminipage\hfill
\vskip -0.7cm
\caption{B-mode APS of the CMB at 10 GHz and 20 GHz induced by the Faraday rotation field with PMF 
normalization comoving scale $\lambda = 1$ Mpc and $n_{B}=0$
(solid black line -- adapted from Fig. 2 in \citealt{kosowskyetal2005}) compared with SKA2 sensitivity (red dashes) achieved in $\sim 10^{3}$ hours of integration on a suitable number of FoVs, 
each of area $\simeq 0.49 \times (1.67 \, {\rm GHz} / \nu)^{2}$ deg$^2$. 
Cosmic+sampling variance from this signal (green three-dots) and instrumental noise limitation (green dots) are also separately displayed. A 10\% binning in $\ell$ is assumed. 
With relatively short baselines exploited here, the sharing of the same integration time on a number of FoVs may be more advantageous in terms of trade-off between the minimization 
of sampling and noise variances.
The use of a focal-plane array with a number of receivers, $N_{r}$, allowing to observe a correspondingly larger sky area in the same time, will imply a better signal-to-noise ratio. See also the text.}
\label{fig:faraday}
\end{figure}

\section{Primordial magnetic fields versus seed magnetic field} 
\label{sec:PMF}

\vskip -0.3cm
The large-scale magnetic fields of the order of few $\mu$G observed in galaxies and galaxy clusters may be the product of the amplification, during structure formation, 
of primordial magnetic seeds (\citealt{2012SSRv..166....1R}). Several early-universe models predict the generation of primordial magnetic fields (PMF),
either during inflation or during cosmological phase transitions (see \citealt{2012SSRv..166...37W} for a review).
The impact of PMF on Big Bang nucleosynthesis constrains their
amplitude at the $\mu$G level (\citealt{2012PhRvD..86f3003K}).  Tighter
constraints at the nG level come from CMB temperature APS
(\citealt{2011PhRvD..83l3533P}, \citealt{2012PhRvD..86d3510S}, \citealt{2013PhLB..726...45P}) and
bispectrum
(\citealt{2009JCAP...06..021C}, \citealt{2009PhRvL.103h1303S}, \citealt{2010PhRvD..82l3006T}, \citealt{2012PhRvL.108w1301T}, \citealt{2012JCAP...03..041S}).
PMF also impact structure formation. The presence of an extra
component of anisotropic stress carried by the PMF and the Lorentz
force induced on baryons modifies the evolution of matter
perturbations, and impacts the small-scale matter power spectrum, and
the formation and early evolution of structure.  This effect, studied
through magnetohydrodynamic $N$-body numerical simulations
(\citealt{2006AN....327..575D}), is still at an early stage of
development, especially concerning the analytical non-linear
treatments.  Data from the $\gamma$-ray observatory {\it Fermi} have
recently added new intriguing observations in the context of
cosmological magnetic fields which might be interpreted as a lower
bound on the PMF amplitude. The data on $\gamma$-ray cascades from
blazars show a lack of photons, which is compatible with diffuse
extra-galactic magnetic fields in the intracluster medium (voids) with
a lower bounds of the order of $10^{-15}-10^{-16}$\,G
(\citealt{2010Sci...328...73N}, \citealt{2011A&A...529A.144T}, \citealt{2012ApJ...747L..14V}).
If this lower bound on the PMF is be confirmed, the SKA could perform
crucial measurements towards probing the PMF generation mechanism.
Current CMB constraints on the PMF are dominated by temperature anisotropy
accuracy measurements, since PMF impact the high multipoles ($\ell \sim
1500$) without suppression by Silk damping. Recent measurements from
WMAP combined with SPT and {\it Planck} were crucial in disentangling
PMF contributions from high-$\ell$ foreground and secondary
anisotropies (\citealt{2013PhLB..726...45P}). SKA measurements in
temperature and polarization of very high-$\ell$ multipoles could
improve these bounds on the PMF as well as the characterization of
foreground and secondary anisotropies beyond the Silk damping
tail. The PMF contribution to CMB anisotropies is generated either at the
last-scattering surface or by Faraday rotation of the intervening
magnetic fields of the stochastic background with the characteristic
frequency dependence $\propto \nu^{-4}$
(\citealt{1996ApJ...469....1K}). The smoking gun of the Faraday rotation
from a stochastic background of PMF is a $B$-polarization signal at very
high-$\ell$ multipoles, with a peak around $\ell \sim 1.4 \times
10^{4}$ (\citealt{kosowskyetal2005}).  SKA2 (in particular with the bands
at $\sim 10 - 20$ GHz, because of their minor foreground
contamination) can target such signal (see Fig. \ref{fig:faraday}) in
the multipole range $\ell \sim 10^{4} - 1.5\times10^{4}$ for a
magnetic field amplitude $\approx$ nG allowed by the temperature
measurement.  We exploit the flexibility of SKA to identify suitable
conditions for detecting this signal,
searching for configurations able to jointly minimize sampling and noise
variances.  We find that, even assuming a collecting area (and,
correspondingly, receiver numbers) decreased to about 50\%, relatively
short baselines (around 5 km) are better suited to this aim.
While the FoV at $\sim 10$ GHz is in principle large enough for a detection 
(or to improve current constraints on PMF models), 
the implementation of SKA2 with focal-plane arrays, allowing an 
increase the observed sky area in the same integration time, will be extremely useful for this research, particularly at 20 GHz. 
Obviously, the best configuration will be selected according to the allocated time and actual implementation of SKA2.
CMB polarization can be crucial to determine or constrain the nature of
the stochastic background, given the different dependence of the Faraday effect on
$B_\lambda$ 
and the magnetic field power spectrum index $n_B$.



\section{Galactic foregrounds}
\label{sec:GalFore}

\vskip -0.3cm
For many of the above topics the accurate study of Galactic emission
is particularly crucial (see e.g. \citealt{buriganaetal2013}).  The
possibility of mapping intermediate and large scales with the SKA
relies on the ability to merge different FoVs into maps with
appropriate large-scale calibration and matching, possibly in
combination with other radio surveys.  On the other hand, the
relatively bright Galactic radio signal does not require the extreme
sensitivity demanded, for instance, by CMB fluctuation mapping at the
SKA highest frequencies.  SKA1 and SKA2 observations in the radio
domain will allow us to test Galactic synchrotron emission models, 3D
physical models of the Galaxy and the large scale coherent
component of the Galactic magnetic field
(\citealt{sunetal08}, \citealt{sunreich09}, \citealt{sunreich10}, \citealt{fauvetetal11}, \citealt{fauvetetal12}),
based on advanced numerical codes (\citealt{strongmoska98}, \citealt{waelkensetal09})
and including turbulence phenomena (\citealt{cholazarian02}).

For both Galactic science and the treatment of foregrounds in
cosmology, it is also important to improve our understanding of the
{\it anomalous microwave emission} (AME) and of the {\it haze}
component. AME is the recently identified emission component which is
well-correlated with far-IR dust emission.  It is produced by rapidly
spinning small dust grains having an electric dipole moment
(\citealt{drainelazarian98}) and its spectrum is expected to peak in the
range 15--50 GHz.  For the first time {\it Planck} was able to define
the shape of the spectrum on the high frequency side of the emission
peak in a number of dust/molecular/HII regions (\citealt{planckXX}).  In
the frequency range 20--40 GHz AME is typically comparable in
brightness to the free-free for the inner Galactic plane.  SKA2 could
provide precise mapping on the low frequency tail of this emission.

{\it Planck} was also able to identify and characterize the emission
from the Galactic haze at microwave wavelengths (\citealt{hazeplanck}). This
is a distinct component of diffuse Galactic emission, roughly centered
on the Galactic centre, extended to $|b| \sim 35^{\circ}$ in Galactic
latitude and $|l| \sim15^{\circ}$ in longitude.  By combining WMAP and
{\it Planck} data, \citet{hazeplanck} were able to determine the
spectrum of this emission to high accuracy, unhindered by the large
systematic biases present in previous analyses.  The derived spectrum
is consistent with power-law emission with a spectral index of $-2.55
\pm 0.05$, thus excluding free-free emission as the source and instead
favoring hard-spectrum synchrotron radiation from an electron
population with a distribution (number density per energy) $dN/dE \sim
E^{-2.1}$. At Galactic latitudes $|b|<30^{\circ}$, the microwave haze
morphology is consistent with that of the Fermi $\gamma$-ray haze or
bubbles (see also \citealt{carrettietal}), indicating that we have a
multi-wavelength view of a distinct component of our Galaxy. Given
the very hard spectrum and the extended nature of the emission,
it is 
unlikely that the haze electrons result from supernova
shocks in the Galactic disk. Instead, a new mechanism for cosmic-ray
acceleration in the centre of our Galaxy is implied.  With the SKA
multifrequency mapping in total intensity and polarization we will
have the opportunity to firmly constrain these models.

In general, a large sky coverage is crucial for mapping Galactic radio emissions.
Among the SKA continuum surveys (\citealt{PrandoniSeymour}), we compare the sensitivity (on the same resolution element) 
of the $\simeq 75$\% sky coverage surveys at 1.4 GHz and at 0.12 GHz, planned for
1--2 years of integration and dedicated respectively to strong gravitational lensing and legacy/rare serendipity and 
to non-thermal emission in clusters and filaments,
with that of radio surveys (\citealt{laportaetal2008}) currently
adopted as ancillary maps in CMB experiment analyses.
The former will have a sensitivity about 20 times better than the available all-sky radio survey at 1.4 GHz, the latter 
will have sensitivity about 4 times better than the Haslam map at 408 MHz (\citealt{haslam}), 
thus representing a significant improvement with respect to current ancillary radio maps.

Summarizing, the SKA high resolution maps of the Galactic emission will contribute to a better understanding of the 
Galactic foreground and will provide key astrophysical information for the separation of CMB and the cosmological HI 21 cm emission.


\smallskip
\smallskip
\noindent {\bf Acknowledgements --} It is a pleasure to thank Isabella Prandoni for useful discussions about SKA performance and flexibility. 
We warmly thank the reviewer and the SKA editors for constructive comments.
Partial support by ASI/INAF Agreement 2014-024-R.0 for the
{\it Planck} LFI Activity of Phase E2 is acknowledged.



\bibliographystyle{apj}

\begin{thebibliography}{plainnat}


\bibitem[Abdalla \& Rawling, 2005]
{Abdalla} \vskip -0.3cm  Abdalla, F.B, Rawling, S., 2005, MNRAS, 360, 27 

\bibitem[Afshordi et al., 2004]
{Afshordi} \vskip -0.3cm  Afshordi, N., Loh, Y.S., Strauss, M.A., 2004, Phys. Rev. D, 69, 083524


\bibitem
[Babich et al., 2004]
{Babich04} \vskip -0.3cm  Babich, D., Creminelli, P., Zaldarriaga, M., 2004,
JCAP, 08, article id. 009

\bibitem[Bartlett \& Silk, 1990]
{BartlettSilk1990} \vskip -0.3cm Bartlett, J.G., Silk, J., 1990, ApJ, 353, 399

\bibitem
[Bartlett \& Stebbins, 1991]
{bart_stebb_1991} \vskip -0.3cm 
Bartlett, J.G., Stebbins, A., 1991, ApJ, 371, 8

\bibitem
[Bartolo et al., 2004]
{Bartolo04} \vskip -0.3cm  Bartolo, N., Komatsu, E., Matarrese, S., Riotto, A., 2010,
Physics Report, 402, 103

\bibitem[Boughn \& Crittenden, 2004a]
{Boughn04a} \vskip -0.3cm  Boughn, S., Crittenden, R., 2004a, Nature, 427, 45

\bibitem[Boughn \& Crittenden, 2004b]
{Boughn04b} \vskip -0.3cm  Boughn, S.P., Crittenden, R.G., 2004b, ApJ, 612, L64

\bibitem[Braun, 2014]{braun2014} \vskip -0.3cm Braun, R., 2014, ``SKA1 Imaging Science Performance``, 
SKA-TEL-SKO-DD-XXX, Rev. A Draft 2

\bibitem
[Burigana et al., 1991]
{BDD91a} \vskip -0.3cm 
Burigana, C., Danese, L., de Zotti, G., 1991, A\&A, 246, 59

\bibitem[Burigana et al., 2013]
{buriganaetal2013} \vskip -0.3cm 
Burigana, C., Davies, R.D., de Bernardis, P., et al., 2013, Int. J. Modern Phys. D, 22, 
1330011

\bibitem
[Burigana et al., 1991]
{BDD95} \vskip -0.3cm 
Burigana, C., de Zotti, G., Danese, L., 1995, A\&A, 303, 323

\bibitem
[Burigana et al., 2004]
{buriganaetalSKA2004} \vskip -0.3cm
Burigana, C., de Zotti, G., Feretti, L., 2004, New Astronomy Reviews, 48, 1107

\bibitem
[Burigana et al., 2008]
{buriganaetal08} \vskip -0.3cm  
Burigana, C., Popa, L.A., Salvaterra, R., et al.,
2008, MNRAS, 385, 404

\bibitem
[Burigana \& Salvaterra, 2003]
{BS03a} \vskip -0.3cm 
Burigana, C., Salvaterra, R., 2003, MNRAS, 342, 543


\bibitem[Camera et al., 2014]
{cameraetal2014} \vskip -0.3cm Camera, S., Santos, M.G., Maartens, R., 2014, arXiv:1409.8286v1

\bibitem[Caprini et al., 2009]
{2009JCAP...06..021C} \vskip -0.3cm  Caprini, C., Finelli, F., Paoletti, D., Riotto, A., 2009, JCAP, 06, article id. 21 

\bibitem[Carretti et al., 2012]
{carrettietal}  \vskip -0.3cm 
Carretti, E., Crocker, R.M., Staveley-Smith, L., et al., 2012, Nature, 493, 66

\bibitem[Carr et al., 2010]
{carretal2010} \vskip -0.3cm Carr, B., Kohri,K., Sendouda, Y., Yokoyama, J., 2010, Phys. Rev. D, 81, 104019

\bibitem[Chluba et al., 2012]
{chlubaal12} \vskip -0.3cm Chluba, J., Erickcek, A.L., Ben-Dayan, I., 2012, ApJ, 758, 76

\bibitem[Chluba  \& Sunyaev, 2012]
{chlubasunyaev2012} \vskip -0.3cm  Chluba, J., Sunyaev, R.A., 2012, MNRAS, 419, 1294

\bibitem[Cho \& Lazarian, 2002]
{cholazarian02} \vskip -0.3cm 
Cho, J., Lazarian, A., 2002, ApJ, 575, L63

\bibitem
[Condon et al., 2012]
{condonetal2012} \vskip -0.3cm 
Condon, J.J., Cotton, W.D., Fomalont, E.B., et al., 2012, ApJ, 758, article id. 23

\bibitem
[COrE Collaboration, 2011]
{corecoll} \vskip -0.3cm 
COrE Collaboration, 2011, ``COrE -- Cosmic Origins Explorer -- A White Paper``, arXiv:1102.2181

\bibitem[Crittenden \& Turok, 1996]
{Crittenden} \vskip -0.3cm  Crittenden, R.G., Turok, N., 1996, PRL, 76, 575 


\bibitem[Danese \& Burigana, 1994]
{daneseburigana94} \vskip -0.3cm Danese, L., Burigana, C., 1994, Lect. Notes Phys., 429, 28

\bibitem
[Danese \& de Zotti, 1980]
{DD80} \vskip -0.3cm 
Danese, L., de Zotti, G., 1980, A\&A, 84, 364

\bibitem[Dewdney et al., 2013]{dewdneyetal2013} \vskip -0.2cm  Dewdney, P., Turner, W., Millenaar, R., 
et al., 2013, ``SKA1 System Baseline Design``, 
SKA-TEL-SKO-DD-001, Rev. 1

\bibitem[Dolag, 2006]
{2006AN....327..575D} \vskip -0.3cm  Dolag, K., 2006, Astronomische Nachrichten, 327, 575 

\bibitem[Draine \& Lazarian, 1998]
{drainelazarian98} \vskip -0.3cm 
Draine, B.T., Lazarian, A., 1998, ApJ, 494, L19


\bibitem[Ejlli \& Dolgov, 2014]
{ejllidolgov2014} \vskip -0.3cm Ejlli, D., Dolgov, A.D., 2014, Phys. Rev. D, 90, 063514


\bibitem[Fauvet et al., 2011]
{fauvetetal11} \vskip -0.3cm 
Fauvet, L., Macias-Perez, J.F., Aumont, J., et al., 2011, A\&A, 526, A145

\bibitem[Fauvet et al., 2012]
{fauvetetal12} \vskip -0.3cm 
Fauvet, L., Macias-Perez, J.F., D\'esert, F.X., 2012, Astroparticle Physics, 36, 57

\bibitem
[Fixsen et al., 1996]
{fixsen96} \vskip -0.3cm 
Fixsen, D.J., Cheng, E.S., Gales, J.M., et al., 1996, ApJ, 473, 576

\bibitem
[Fixsen \& Mather, 2002]
{FM02} \vskip -0.3cm 
Fixsen, D.J., Mather, J.C., 2002, ApJ, 581, 817

\bibitem[Fosalba et al., 2003]
{Fosalba03} \vskip -0.3cm  Fosalba, P., Gazta÷naga, E., Castander, F.J., 2003, ApJ, 597, L89  

\bibitem[Fosalba \& Gaztanaga, 2004]
{Fosalba04} \vskip -0.3cm  Fosalba, P., Gaztanaga, E., 2004, MNRAS, 350, L37 

\bibitem[Francis \& Peacock, 2010]
{Francis} \vskip -0.3cm  Francis, C.L., Peacock, J.A., 2010, MNRAS, 406, 2 


\bibitem[Gervasi et al., 2008]{gervasietal2008} \vskip -0.3cm Gervasi, M., Tartari, A., Zannoni, M., Boella, G., Sironi, G., 2008, ApJ, 682, 223

\bibitem[Giannantonio et al., 2014]
{Giannantonio_etal} \vskip -0.3cm Giannantonio, T., Ross, A.J., Percival, W.J.,
2014, Phys. Rev. D, 89, 
023511

\bibitem[Granett et al., 2008]
{Granett08} \vskip -0.3cm  Granett, B.R., Neyrinck, M.C., Szapudi, I., 2008, ApJ, 683, L99 

\bibitem[Granett et al., 2009]
{Granett09} \vskip -0.3cm  Granett, B.R., Neyrinck, M.C., Szapudi, I., 2009, ApJ, 701, 414 


\bibitem[Haslam et al., 1982]{haslam} \vskip -0.3cm Haslam, C.G.T., Salter, C.J., Stoffel, H., Wilson, W.E. 1982, A\&AS, 47, 1



\bibitem[Jedamzik et al., 2000]
{jedamziketal2000} \vskip -0.3cm Jedamzik, K.,Katalinic V., Olinto, A.V., 2000, PRL, 85, 700


\bibitem[Kawasaki \& Kusakabe, 2012]
{2012PhRvD..86f3003K} \vskip -0.3cm  Kawasaki, M., Kusakabe, M., 2012, Phys. Rev. D, 86, 063003 

\bibitem
[Khatri \& Sunyaev, 2012]
{khatrisunyaev} \vskip -0.3cm 
Khatri, R., Sunyaev, R.A., 2012, JCAP, 09, article id. 016

\bibitem
[Kogut, 1996]
{KOG96} \vskip -0.3cm 
Kogut, A., 1996, 
in XVI Moriond Astrophysics meeting, 
March 16-23, Les Arcs, France,
astro-ph/9607100

\bibitem
[Kogut, 2003]
{KOG03} \vskip -0.3cm 
Kogut, A., 2003,
New Astronomy Reviews, 
47, 
945

\bibitem
[Kogut et al., 2011]
{kogutpixie} \vskip -0.3cm 
Kogut, A., Fixsen, D.J., Chuss, D.T., et al., 2011, JCAP, 07, article id. 025 

\bibitem
[Komatsu et al., 2010]
{Komatsu10} \vskip -0.3cm  Komatsu, E., 
 Afshordi, N., Bartolo, N., et al., 2010, 
 in {Astro2010: The Astronomy and Astrophysics Decadal Survey}, 
 Science White Papers, Vol. 158,  arXiv:0902.4759

\bibitem
[Komatsu \& Spergel, 2001]
{Komatsu01} \vskip -0.3cm  Komatsu, E., Spergel, D.N., 2001,
Phys. Rev. D, 63, 063002

\bibitem[Kosowsky et al., 2005]
{kosowskyetal2005} \vskip -0.3cm  Kosowsky, A., Kahniashvili, T., Lavrelashvili, G., Ratra, B., 2005, Phys. Rev. D, 71, 043006

\bibitem[Kosowsky \& Loeb, 1996]
{1996ApJ...469....1K} \vskip -0.3cm  Kosowsky, A., Loeb, A., 1996, ApJ, 469, 1

\bibitem
[Kurz \& Shaver, 1999]
{kurzshaver1999} \vskip -0.3cm  Kurz, R., Shaver, P., 1999, Messenger, 96, 7


\bibitem[La Porta et al., 2008]{laportaetal2008} \vskip -0.3cm La Porta, L., Burigana, C., Reich, W., Reich, P., 2008, A\&A, 479, 641

\bibitem[Leisted et al., 2014]
{Leistedt_etal2014} \vskip -0.3cm Leistedt, B, Hiranya, H.V., Roth, N., 2014, arXiv:1405.4315v3

\bibitem[LSST DE Science Collaboration, 2012]
{LSSTDEColl}  \vskip -0.3cm LSST Dark Energy Science Collaboration, 2012, arXiv:1211.0310


\bibitem
[Mather et al., 1990]
{mather90} \vskip -0.3cm 
Mather, J.C., Cheng, E.S., Eplee, R.E. Jr.,
et al., 1990, ApJ, 354, L37

\bibitem[Mauch et al., 2013]
{mauchetal2013} \vskip -0.3cm  Mauch, T., Kl\"ockner, H.R., Rawlings, S., et al., 2013, MNRAS, 435, 650


\bibitem
[Naselsky	\& Chiang, 2004]
{naselskychiang04} \vskip -0.3cm 
Naselsky,	P., Chiang, L.Y., 2004, MNRAS, 347, 921

\bibitem[Neronov \& Vovk, 2010]
{2010Sci...328...73N} \vskip -0.3cm  Neronov, A., Vovk, I., 2010, Science, 328, 73 

\bibitem[Nolta et al., 2004]
{Nolta} \vskip -0.3cm  Nolta, M.R.,Wright, E.L., Page, L., et al., 2004, ApJ, 608, L10 


\bibitem
[Oh, 1999]
{OH99} \vskip -0.3cm 
Oh, S.P., 1999, ApJ, 527, 16

\bibitem[Ostriker \& Thompson, 1987]
{ostrikerthompson87} \vskip -0.3cm Ostriker, J.P., Thompson, C., ApJ, 323, L97


\bibitem[Padmanabhan et al., 2005]
{Padmanabhan} \vskip -0.3cm  Padmanabhan, N., Hirata, C.M., Seljak, U., et al., 2005, Phys. Rev. D, 72, 043525 

\bibitem[Pani \& Loeb, 2013]
{paniloeb2013} \vskip -0.3cm Pani, P., Loeb, A., 2013, Phys. Rev. D, 88, 041301

\bibitem[Paoletti \& Finelli, 2011]
{2011PhRvD..83l3533P} \vskip -0.3cm  Paoletti, D., Finelli, F.,  2011, Phys. Rev. D, 83, 123533 

\bibitem[Paoletti \& Finelli, 2013]
{2013PhLB..726...45P} \vskip -0.3cm  Paoletti, D., Finelli, F., 2013, Phys. Lett. B, 726, 45 

\bibitem[Papai et al., 2011]
{Papai} \vskip -0.3cm  Papai, P., Szapudi, I., Granett, B.R., 2011, ApJ, 732, L27 

\bibitem
[Partridge et al., 1997]
{Partridgeetal1997} \vskip -0.3cm 
Partridge, R.B., Richards, E.A., Fomalont, E.B., Kellerman, K.I., Windhorst, R., 1997, ApJ, 483, 38

\bibitem[Peiris \& Spergel, 2000]
{Peiris} \vskip -0.3cm  Peiris, H.V., Spergel, D.N., 2000, ApJ, 540, 605 

\bibitem[{\it Planck} Collaboration, 2011]
{planckXX}  \vskip -0.3cm 
{\it Planck} Collaboration, 
2011, A\&A, 536, A20

\bibitem[{\it Planck} Collaboration, 2013]
{hazeplanck} \vskip -0.3cm 
{\it Planck} Collaboration, 
2013, A\&A, 554, A139 

\bibitem
[{\it Planck} Collaboration, 2014a]
{planck19} \vskip -0.3cm 
{\it Planck} Collaboration, 2014a, {\it Planck} 2013 results. XIX,
A\&A, 571, A18 

\bibitem
[{\it Planck} Collaboration, 2014b]
{PlanckCollNonGauss} \vskip -0.3cm 
{\it Planck} Collaboration, 2014b, {\it Planck} 2013 results. XXIV, 
 A\&A, 571, A24


\bibitem
[Ponente et al., 2011]
{ponenteetal2011} \vskip -0.3cm 
Ponente, P.P., Diego, J.M., Sheth, R.K., et al.,
2011, MNRAS, 410, 2353

\bibitem
[Prandoni et al., 2001]
{prandonietal01} \vskip -0.3cm 
Prandoni, I., Gregorini, L., Parma, P., et al., 2001, A\&A, 365, 392

\bibitem[Prandoni \& Seymour, 2014]
{PrandoniSeymour} \vskip -0.3cm  Prandoni, I., Seymour, N., 2014, PoS, {This Issue}

\bibitem
[PRISM Collaboration, 2014]
{prismjcap} \vskip -0.3cm 
PRISM Collaboration, 2014, JCAP, 02, article id. 006

\bibitem
[Procopio \& Burigana, 2009]
{procopioburigana} \vskip -0.3cm 
Procopio, P., Burigana, C., 2009, A\&A, 507, 1243



\bibitem[Raccanelli et al., 2008]
{Raccanelli} \vskip -0.3cm  Raccanelli, A., Bonaldi, A., Negrello, M., et al., 2008, MNRAS, 386, 2161 

\bibitem[Raccanelli et al., 2014]
{raccanellietal2014} \vskip -0.3cm Raccanelli, A., Dor\'e, O., David J. Bacon, D.J., et al., 2014, arXiv:1406.0010v1

\bibitem[Rassat et al., 2007]
{Rassat} \vskip -0.3cm  Rassat, A., Land, K., Lahav, O., Abdalla, F.B., 2007, MNRAS, 377, 1085 

\bibitem[Ryu et al., 2012]
{2012SSRv..166....1R} \vskip -0.3cm  Ryu, D., Schleicher, D.R.G., Treumann, R.A., Tsagas, C.G., Widrow, L.M., 2012, Space Sci. Rev., 166, 1 


\bibitem[Sachs \& Wolfe, 1967]
{SachsWolfe} \vskip -0.3cm  Sachs, R.K., Wolfe, A.M., 1967, ApJ, 147, 73

\bibitem[Salvaterra \& Burigana, 2002]{SB02} \vskip -0.3cm Salvaterra, R., Burigana, C., 2002, MNRAS, 336, 592

\bibitem[Schiavon et al., 2012]
{schiavonetal2012} \vskip -0.3cm 
Schiavon, F., Finelli, F., Gruppuso, A., et al.,
2012, MNRAS, 427, 3044

\bibitem
[Schneider et al., 2008]
{Schneideretal2008} \vskip -0.3cm 
Schneider, R., Salvaterra, R., Choudhury, T.R., et al.,
2008, MNRAS, 384, 1525

\bibitem
[Seiffert et al., 2011]
{seiffertetal2011} \vskip -0.3cm 
Seiffert, M.D., Fixsen, D.J., Kogut, A., et al., 2011, ApJ, 734, article id. 6

\bibitem[Seshadri \& Subramanian, 2009]
{2009PhRvL.103h1303S} \vskip -0.3cm  Seshadri, T.R., Subramanian, K., 2009, PRL, 103, 081303 

\bibitem[Shaw \& Lewis, 2012]
{2012PhRvD..86d3510S} \vskip -0.3cm  Shaw, J.R., Lewis, A., 2012, Phys. Rev. D, 86, 043510 

\bibitem[Shiraishi et al., 2012]
{2012JCAP...03..041S} \vskip -0.3cm  Shiraishi, M., Nitta, D., Yokoyama, S., Ichiki, K., 2012, JCAP, 03, article id. 41 

\bibitem
[Singal et al., 2011]
{singaletal2011} \vskip -0.3cm 
Singal, J., Fixsen, D., Kogut, A., et al., 2011, ApJ, 730, article id. 138

\bibitem
[Singal et al., 2010]
{singaletalMN} \vskip -0.3cm 
Singal, J., Stawarz, L., Lawrence, A., Petrosian, V., 2010, MNRAS, 409, 1172

\bibitem
[Stebbins \& Silk, 1986]
{stebb_silk} \vskip -0.3cm 
Stebbins, A., Silk, J., 1986, ApJ, 169, 1

\bibitem[Strong \& Moskalenko, 1998]
{strongmoska98} \vskip -0.3cm 
Strong, A.W., Moskalenko, I.V., 1998, ApJ, 509, 212

\bibitem[Sun et al., 2008]
{sunetal08} \vskip -0.3cm 
Sun, X.H., Reich, W., Waelkens, A., Ensslin, T.A., 2008, A\&A, 477, 573

\bibitem[Sun \& Reich, 2009]
{sunreich09} \vskip -0.3cm 
Sun, X.H., Reich, W., 2009, A\&A, 507, 1087

\bibitem[Sun \& Reich, 2010]
{sunreich10} \vskip -0.3cm 
Sun, X.H., Reich, W., 2010, Research in Astron. Astrophys., 10, 1287

\bibitem[Sunyaev \& Khatri, 2013]
{sunyaevkhatri2013}  \vskip -0.3cm Sunyaev, R.A., Khatri, R., 2013, Int. J. Modern Phys. D, 22, 1330014

\bibitem
[Sunyaev \& Zeldovich, 1970]
{SZ70} \vskip -0.3cm 
Sunyaev, R.A., Zeldovich, Ya.B., 1970, Ap\&SS, 7, 20


\bibitem[Takahashi et al., 2014]{takahashi etal2014thisissue} \vskip -0.3cm Takahashi, K., Brown, M.L., Burigana, C., et al., 
2014, PoS, This issue

\bibitem[Taylor et al., 2011]
{2011A&A...529A.144T} \vskip -0.3cm  Taylor, A.M., Vovk, I., Neronov, A., 2011, A\&A, 529, A144 

\bibitem[Tegmark, 1997]
{Tegmark1997} \vskip -0.3cm 
Tegmark, M., 1997, Phys. Rev. D, 55, 5895

\bibitem[Trivedi et al., 2012]
{2012PhRvL.108w1301T} \vskip -0.3cm  Trivedi, P., Seshadri, T.R., Subramanian, K., 2012, PRL, 108, 231301 

\bibitem[Trivedi et al., 2010]
{2010PhRvD..82l3006T} \vskip -0.3cm  Trivedi, P., Subramanian, K., Seshadri, T.R., 2010, Phys. Rev. D, 82, 123006 

\bibitem
[Trombetti \& Burigana, 2012]
{trombettiburigana12} \vskip -0.3cm 
Trombetti, T., Burigana, C., 2012, J. Modern Phys., 3, 1918

\bibitem
[Trombetti \& Burigana, 2014]
{trombettiburigana14} \vskip -0.3cm 
Trombetti, T., Burigana, C., 2014, MNRAS, 437, 2507



\bibitem
[Verde et al., 2000]
{Verde00} \vskip -0.3cm  Verde, L., Wang, L.M., Heavens, A., Kamionkowski, M., 2000,
MNRAS, 313, L141

\bibitem[Vernstrom et al., 2014a]
{vernstrometal2014}  \vskip -0.3cm Vernstrom, T., Norris, Ray P., Scott, D., Wall, J.V., 2014a, arXiv:1408.4160v1

\bibitem[Vernstrom et al., 2014b]
{vernstrometal2014mnras} \vskip -0.3cm  Vernstrom, T., Scott, D., Wall, J.V., et al.,
2014b, MNRAS, 404, 2791

\bibitem[Vovk et al., 2012]
{2012ApJ...747L..14V} \vskip -0.3cm  Vovk, I., Taylor, A.M., Semikoz, D., Neronov, A., 2012, ApJ, 747, L14 


\bibitem[Waelkens et al., 2009]
{waelkensetal09} \vskip -0.3cm 
Waelkens, A., Jaffe, T., Reinecke, M., Kitaura, F.S., Ensslin, T.A., et al., 2009, A\&A, 495, 697

\bibitem[Widrow et al., 2012]
{2012SSRv..166...37W} \vskip -0.3cm  Widrow, L.M., Ryu, D., Schleicher, D.R.G., et al., 2012, Space Sci. Rev., 166, 37 

\bibitem[Williams et al., 2014]
{williamsetal2013} \vskip -0.3cm  Williams, W.L., Intema, H.T., R\"ottgering, H.J.A., 2013, A\&A, 549, A55


\bibitem[Xia et al., 2011]
{Xia11} \vskip -0.3cm  Xia, J.Q., Baccigalupi, C., Matarrese, S., Verde, L., Viel, M., 2011, JCAP, 08, article id. 033

\bibitem[Xia et al., 2010a]
{Xia10b} \vskip -0.3cm  Xia, J.Q., Bonaldi, A., Baccigalupi, C., 
2010a, JCAP, 08, article id. 013

\bibitem[Xia et al., 2009]
{Xia09} \vskip -0.3cm  Xia, J.Q., Viel, M., Baccigalupi, C., Matarrese, S., 2009, JCAP, 09, article id. 003

\bibitem[Xia et al., 2010b]
{Xia10a} \vskip -0.3cm  Xia, J.Q., Viel, M., Baccigalupi, C., et al.,
2010b, ApJ, 717, L17


\end{thebibliography}

\end{document}